\documentclass[aps,pra,showpacs,preprint,superscriptaddress]{revtex4}
\usepackage{amsmath,amssymb,amsfonts}
\usepackage[english]{babel}
\usepackage{graphicx}
\usepackage{dcolumn}
\usepackage{bm}
\usepackage{color}

\begin{document}

\title{A single structured light beam as an atomic cloud splitter}

\author{B. M. Rodr\'{\i}guez-Lara}
\email[]{bmlara@fisica.unam.mx}
\affiliation{Instituto de F\'{\i}sica, Universidad Nacional Aut\'{o}noma de M\'{e}xico,
Apdo. Postal 20-364, M\'{e}xico D.F. 01000, M\'{e}xico.}

\author{R. J\'auregui}
\email[]{rocio@fisica.unam.mx}
\affiliation{Instituto de F\'{\i}sica, Universidad Nacional Aut\'{o}noma de M\'{e}xico,
Apdo. Postal 20-364, M\'{e}xico D.F. 01000, M\'{e}xico.}

\date{\today}

\begin{abstract}
We propose a scheme to split a cloud of cold non-interacting neutral atoms based
on their dipole interaction with a single structured light beam which exhibits
parabolic cylindrical symmetry. Using semiclassical numerical simulations, we establish
a direct relationship between the general properties of the light beam and the relevant
geometric and kinematic properties acquired by the atomic cloud as its passes through
the beam.
\end{abstract}

\maketitle

An optimal implementation of an atomic thermal cloud splitter requires achievement
of a predetermined efficiency and deviation angle, as well as prevention of heating
processes. Optical devices for splitting a free falling atomic cloud usually rely on
two, or more, non-collinear \cite{Houde2000} or collinear \cite{Bragg1988} light beams,
since a single Gaussian laser beam orthogonal to the falling atoms yields deviations
of only some ten milliradians \cite{single}. On chip, optical splitting requires either
a pulse sequence of tightly linearly polarized beams \cite{Wang2005} or the splitting
into two of a former one well dipole trapping beam via acousto-optical modulation \cite{Pasquini2005}.
It is also possible to use on chip magnetic fields to split the atomic cloud by
including current carrier wires with varying bias or distance between them \cite{Kraft2005, Hommelhoff2005}.

Here, we propose to reconsider the simple configuration of a single but {\it structured}
laser beam orthogonal to a free falling atomic thermal cloud. Light beams with special intensity
or phase structure yielding peculiar dynamical properties have been extensively used for
the manipulation of microparticles \cite{Dholakia} and cold atoms \cite{OAM_BEC}. Most
theoretical and experimental studies for cold atoms focus on using such light beams as
wave guides for atom transport with the possibility of exchanging orbital angular momentum
\cite{OAM_BEC, allen92, TabosaP99,luciana, bec current,nienhuis,twist,loops08} or other
dynamical properties \cite{blas08} between the light and the atoms. We shall show that
experimentally accessible structured light beams can also be used efficiently as atomic
thermal cloud splitters. This proposal is based on semiclassical calculations that allow an estimate
of the relevant geometrical and mechanical properties of the atomic cloud, such as deviation angles
and kinetic energy, in terms of the  atomic and light beams setup. In these calculations,
the center of mass motion of each atom is described using Newtonian mechanics with the optical force derived from quantum theory \cite{Gordon1980}.

\begin{figure}[ht!]
\includegraphics[width= 0.9 \textwidth]{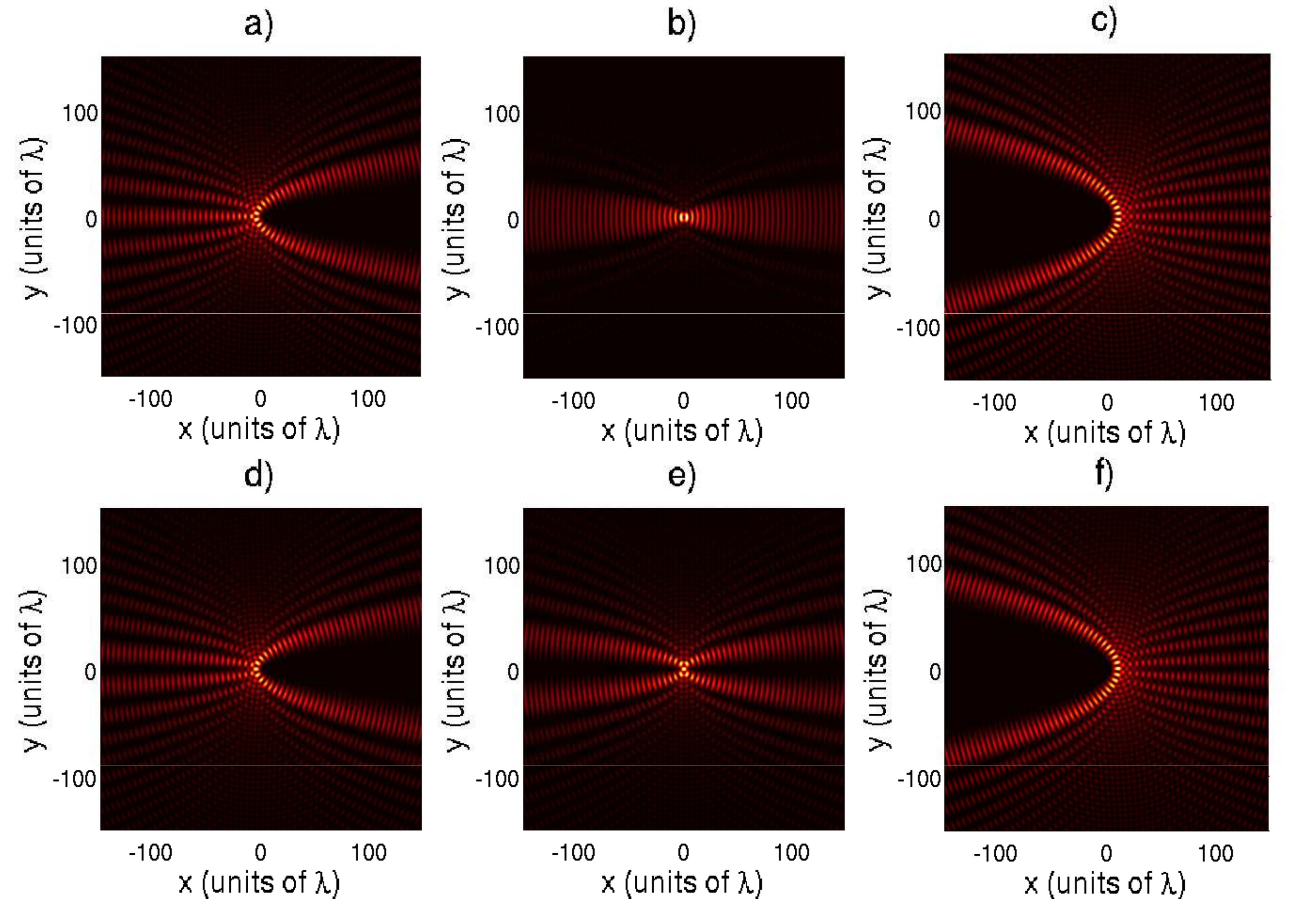}
\caption{\label{fig:1}(Color Online) Sampler of the transverse structure of the intensity of
Weber EM fields as defined by
Eqs.~(\ref{eq:total}). Even/odd EM Weber fields with eigenvalues (a/d) $a=-2$,
(b/e) $a=0$,  (c/f) $a=5$.}
\end{figure}

\begin{figure}[ht!]
\includegraphics[width= 0.8 \textwidth]{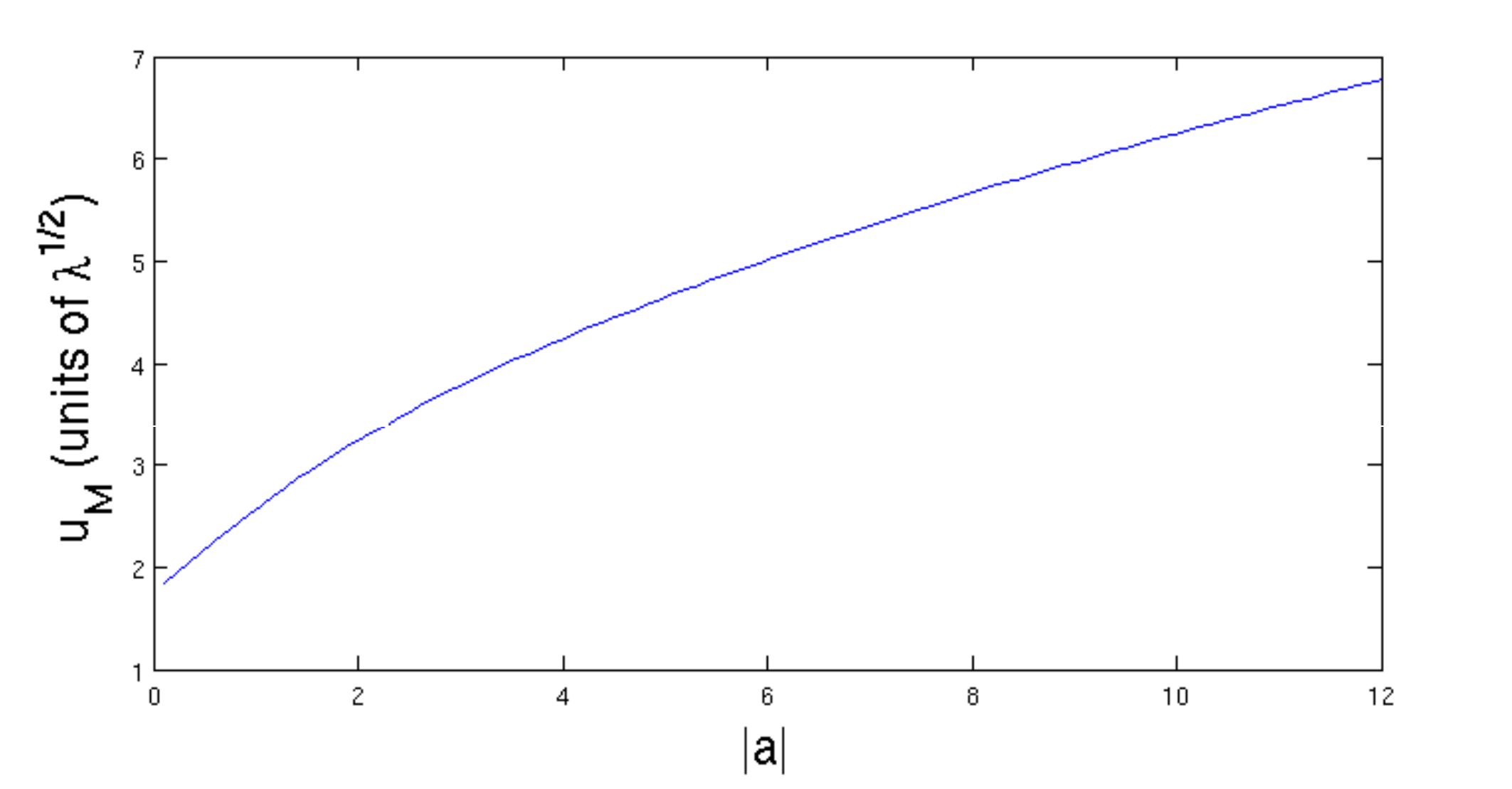}
\caption{\label{fig:2}(Color Online) Value for $u_M$, the coordinate that defines the parabola delimiting the low intensity region of a Weber field, as a function of the eigenvalue $a$. The length unit was taken as the wavelength of the beam and $k_z = 0.995 k$.}
\end{figure}

The structured light beams we study are known as vector Weber beams.
Under ideal conditions, the intensity pattern
of an electromagnetic (EM) Weber beam is invariant along the main direction of propagation.
Its transverse structure is naturally described in terms of parabolic coordinates $u$
and $v$ related to the Cartesian coordinates through the equations $x =(u^2-v^2)/2$,
$y=uv$, $u\epsilon(-\infty,\infty)$, $v\epsilon[0,\infty)$. Its associated electric and
magnetic fields can be written as a superposition of transverse electric (TE) and magnetic
(TM) modes \cite{Stratton1941}
\begin{eqnarray}
\bm{E}_\kappa &=& - \partial_{ct}  \left( \mathcal{A}^{(TE)} \bm{\mathbb{M}}  - \mathcal{A}^{(TM)} \bm{\mathbb{N}} \right) \Psi_\kappa, \nonumber \\
\bm{B}_\kappa &=& \quad\partial_{ct}  \left( \mathcal{A}^{(TE)} \bm{\mathbb{N}}  - \mathcal{A}^{(TM)} \bm{\mathbb{M}} \right) \Psi_\kappa. \label{eq:total}
\end{eqnarray}
The vector operators $\bm{\mathbb{M}}$ and $\bm{\mathbb{N}}$ are
 \begin{eqnarray}
\bm{\mathbb{M}} &=& h^{-1}\partial_{ct} \left( \bm{e}_{u} \partial_{v} - \bm{e}_{v}\partial_{u} \right), \nonumber\\
\bm{\mathbb{N}} &=& h^{-1}\partial_{z}  \left( \bm{e}_{u} \partial_{u} + \bm{e}_{v} \partial_{v} \right) - \bm{e}_{z} \nabla_{\perp}^2,
\end{eqnarray}
where $h =\sqrt{u^2 + v^2} $ denotes the scaling factor, $\bm{e}_{\zeta}$, $\zeta =u$, $v$, $z$,
the unitary vectors corresponding to parabolic-cylindrical
coordinates, and  the shorthand notation $\partial_{x}$ is
used for partial derivatives with respect to the variable $x$. The scalar field $\Psi_{\kappa}$ is the solution
of the wave equation:
\begin{eqnarray}
&&\Psi_{\kappa}(u,v,z,t)= \psi_{\mathfrak{p},k_\perp,a}(u,v)e^{i \left( k_{z} z -\omega t \right)}\nonumber\\
&=& \tilde U_{\mathfrak{p},k_\perp,a}(u) \tilde V_{\mathfrak{p},k_\perp,a}(v) e^{-i\left(\frac{ k_{\perp} (u^{2}+v^2)}{  2}+ k_{z} z -\omega t \right)},
\label{eq:phiscalar}\end{eqnarray}
where
\begin{eqnarray}
\tilde U_{\mathfrak{p},k_\perp,a}(u) &=& (k_{\perp} u^2)^{\frac{n_{\mathfrak{p}}-1}{4}}~_{1}F_{1} \left( \frac{n_{\mathfrak{p}}}{4}- i
\frac{a}{2}, \frac{n_{\mathfrak{p}}}{2}; i k_{\perp} u^2 \right),\nonumber\\
\tilde V_{\mathfrak{p},k_\perp,a}(v) &=& (k_{\perp} u^2)^{\frac{
n_{\mathfrak{p}}-1}{4}}~_{1}F_{1} \left( \frac{n_{\mathfrak{p}}}{4}+ i
\frac{a}{2}, \frac{n_{\mathfrak{p}}}{2}; i k_{\perp} u^2 \right),\nonumber
\end{eqnarray}
with $n_{\mathfrak{p}}=1$ ($n_{\mathfrak{p}}=3$ ) for even (odd) parity functions,
 $_{1}F_{1}(x_1,x_2,x_3)$ is the confluent hypergeometric function directly related to Weber functions \cite{Lebedev1972,Abramowitz1972,Volke2006}.
 Up to a normalization factor, the scalar wave function $\psi_{\mathfrak{p},
k_{\perp},a}$ in terms of the Weber beam angular spectra $\mathfrak{A}$ is given by ~\cite{Bandres2004}:
\begin{eqnarray}
\psi_{\mathfrak{p},k_\perp,a}(x,y) &=&\int_{-\pi}^\pi \mathfrak{A}_p(a;\varphi)e^{-ik_\perp(x\cos\varphi + y\sin\varphi)}d\varphi
\\
\mathfrak{A}_e(a;\varphi) &=& \frac{e^{ia\ln\vert\tan\varphi/2\vert}}{2\sqrt{\pi\sin\varphi}}\\
\mathfrak{A}_o(a;\varphi) &=& \left\{ \begin{array}{cl} i \mathfrak{A}_e(a;\varphi)&\varphi\epsilon(-\pi,0)\\ -i\mathfrak{A}_e(a;\varphi)&\varphi\epsilon(0,\pi). \end{array}\right.
\end{eqnarray}
The labels set $\kappa=\{\mathfrak{p},k_z,\omega,a\}$ in Eq.~(\ref{eq:total}) refers to the mathematical parity,
$\mathfrak{p}$ even or odd, $z$-component of the wave vector, $k_z$, frequency, $\omega =c k= c\sqrt{k_{z}^2 + k_{\perp}^2}$,
and  continuous order, $a\in \mathbb{R}$.
The Weber beam transverse structure exhibits extended zones where the field has zero, or close to zero,
values as illustrated in Fig.~(\ref{fig:1}).  For the order parameter $a\ne 0$, the low intensity
zone is to the right (left) depending on the parameter $a$ being negative
(positive); its area increases in size as the absolute value of $a$
increases in magnitude. This dark region is approximately delimited by
the parabola $ y = \sqrt{2 u_{M}^2 (\vert x \vert-u_M^2/2)}$ where $u_{M}$
is the value of $u$ at the first maxima of the function $U_{\mathfrak{p},k_\perp,a}(u)$.
In Fig.~(\ref{fig:2}), $u_M$ as a function of $a$ is illustrated.

The dynamical constants of Weber photons are \cite{us}:
its momentum along $z$, $\hbar k_z$,  its energy $\hbar \omega$, and its eigenvalue, $\hbar^2k_\perp a$, for the quantum operator
  \begin{equation}
  \hat{\mathbb{A}}_{EM}
=\hbar\sum_{i=1}^3 \int dV \hat E_i\frac{(l_zp_y+p_yl_z)}{2}\hat A_i
=\sum_{i,\kappa}  \hbar^2 k_{\perp} a ~\hat{N}_{\kappa}^{(i)}.\label{eq:A}
 \end{equation}
Here $\hat{N}_{\kappa}^{(i)}$ denotes the number operator of the $\kappa$ mode,
$l_z = -i(\vec r\times\vec\nabla)_z$ the generator of rotations along the $z$-axis
and $p_y = -i \partial_y$ the generator of translations along the $y$-axis. Then, it would be said that
the parameter $a$ determines the product of the angular momentum along $z$ and the momentum
along $y$ of a Weber photon whenever it could be shown that charged particles exchange that
mechanical property with those photons.

Recently, Weber beams of zero-order were experimentally generated by means of a thin annular slit modulated by the proper angular spectra \cite{Lopez2005}. This setup was conceived as a variation of that originally used by Durnin et al.\cite{Durnin} for generating Bessel beams. Higher order Weber beams can be produced using holograms encoded either on plates \cite{Lopez2005} or on spatial light modulators \cite{arrizon}.

The interaction between a two level atom and an EM Weber wave with a frequency close to resonance is determined by the coupling factor
\begin{eqnarray}
g^{\pm} &=& \bm{\mu}_{12}^{\pm} \cdot \bm{E}\nonumber\\
&=& \frac{\mu^{\pm} k}{h^2} (u\pm \imath v) (k \mathcal{A}^{(TE)} \mp \imath k_{z} \mathcal{A}^{(TM)})(\partial_{v} \mp \imath \partial_{u} ) \Psi_\kappa,\nonumber
\end{eqnarray}
where the $\pm$ super index refers to the two possible cases for the atomic dipole
transition element $\bm{\mu}_{12}^{\pm} = \mu_{12}^{\pm}  (\bm{e}_{x} \pm \imath \bm{e}_{y})$. In a semiclassical treatment, the gradient of $g$,
${\bm\nabla} g= ({\bm\alpha} +i  {\bm \beta}) g$, defines the force experienced by the atom.
The expression for the average semiclassical velocity dependent force valid for both
propagating and standing beams is \cite{Gordon1980}:
\begin{equation}
\langle {\bm f} \rangle = \frac{\hbar\Gamma\Big[[(
D(1-p){\bm v}\cdot{\bm \alpha}) +\Gamma/2] {\bm  \beta} + [({\bm
v}\cdot {\bm \beta}) -
\delta\omega]{\bm\alpha}\Big]}{(1-p^\prime)p^{\prime -1}\Gamma +2D{\bm v}\cdot{\bm \alpha}[1-p/p^\prime -p]} \label{eq:force}
\end{equation}
with $\Gamma$ the Einstein coefficient, $\Gamma =
4k^3\vert{\bm\mu}_{12}\vert^2/3\hbar$, $\delta\omega$ the detuning
between the wave frequency $\omega$ and the transition frequency
$\omega_0$,  $\delta\omega = \omega -\omega_0$, $p =2\vert
g\vert^2/((\Gamma/2)^2+\delta\omega^2)$ a saturation parameter linked to the
difference $D$ between the populations of the  atom  two levels,
 $D=1/(1+p)$, and finally
$p^\prime = 2\vert g\vert^2/\vert \gamma^\prime\vert^2$, with
$\gamma^\prime =({\bm v}\cdot{\bm\alpha})(1-p)(1+p)^{-1} +\Gamma/2 +
\imath (-\delta\omega +({\bm v}\cdot{\bm\beta}))$.

Our proposed scheme is focused on the  red detuned far off resonance
case so that nonconservative terms arising from the velocity
dependence of the force are not dominant \cite{Miller1993}.
Nevertheless, we keep the velocity dependent terms in the numerical calculations
in order to prevent disregarding
potentially relevant effects, since the atom may increase its kinetic energy
as it interacts with the light beam. The numerical
simulations consider a TE laser beam detuned 67 nm to the red of the $5~^2S_{1/2}$ -
$5~^2P_{1/2}$ transition at $795$~nm of $^{85}\text{Rb}$ with irradiance
in the range $\sim 1.5-6$ W/ cm$^2$. As natural length and time units, we take
the laser wavelength and the inverse of the Einstein coefficient $\Gamma$,
which is $3.7 \times 10^{7}$s$^{-1}$ for the state $5~^2P_{1/2}$ of $^{85}\text{Rb}$.

An atomic cloud is released into a region where a Weber beam propagating in the horizontal
direction was generated. The atoms have an initial velocity
oriented by the gravitational field in the negative $x$-direction, with
a random smaller component in the $y$ or $z$-direction (we shall take
$\vert v_x\vert \geq 10 v_\perp = 10 (v_y^2+v_z^2)^{1/2}$).
The Weber parameter $a$ is taken negative so that, initially, the atomic cloud faces the low intensity region; for $x>0$, due to the red detuning of the light,
the atoms that arrive
in the dark region of the beam are attracted towards the delimiting parabola mentioned above
and the splitting of the cloud takes place. In the region $x<0$ these atoms find a series
 of effective potential well channels that give rise to their focusing towards those
 channels. The atoms that arrive to the Weber beam outside the central dark region
  experience either a strong deflection when their initial position is on the bright parabolas,
 or a weak deflection if their initial position is outside those parabolas.
 This behavior was found to be generic in all numerical simulations we performed.
 For the parameters mentioned above and  $mv_x^2/2k_B \sim 1\mu$K, the irradiance
 can be chosen so that
 the maximum deflection angle for the atoms that arrive in the dark region
 is  $\theta_d^{\max}\sim \arctan ((u_M^2/\vert x_0 \vert(1 -u^2_M/2\vert x_0 \vert ))^{1/2})$, where
 $x_0$ is the initial atom $x$-coordinate and $u_M$ is the coordinate that defines the parabola delimiting the dark region; $u_M$ is determined by the $a$ parameter of the Weber beam, Fig.(\ref{fig:2}).
 As expected, higher (lower) initial atomic kinetic energies require
 higher (lower) irradiance to achieve such deflection angles.
\begin{figure}[ht!]
\includegraphics[width= 0.9 \textwidth]{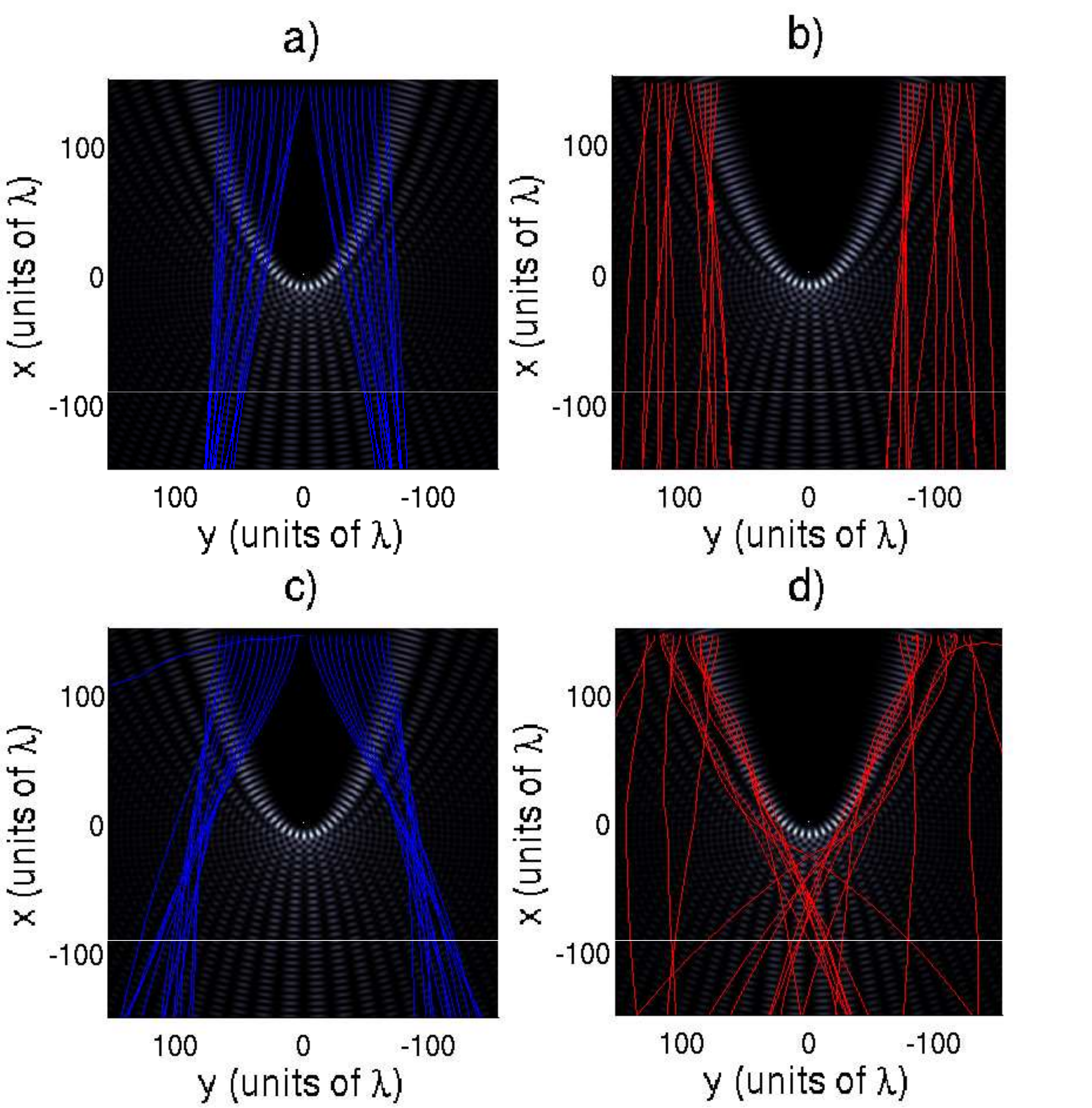}
\caption{\label{fig:Fig3}(Color Online) Evolution of an atomic cloud as it interacts with an odd TE Weber beam whose intensity pattern is illustrated in black and white. The atoms initial  velocities are mainly in the direction of the gravitational field, $v_x \sim -0.6~m\lambda\Gamma$ with random $v_y$ and $v_z$ components, $\vert v_x\vert \ge 10v_\perp$, corresponding to $T \sim 1.5~\mu K$. The Weber beam is characterized by $a=-5$, $k_{z} = 0.995~k$ and irradiance (a,b) $1.725~ W/ cm^2$ and (c,d) $6~ W/ cm^2$. The atoms motion takes place within a $\sim 20 ~\lambda$ wide region in the $z$-axis.}
\end{figure}

\begin{figure}[ht!]
\includegraphics[width= 0.9 \textwidth]{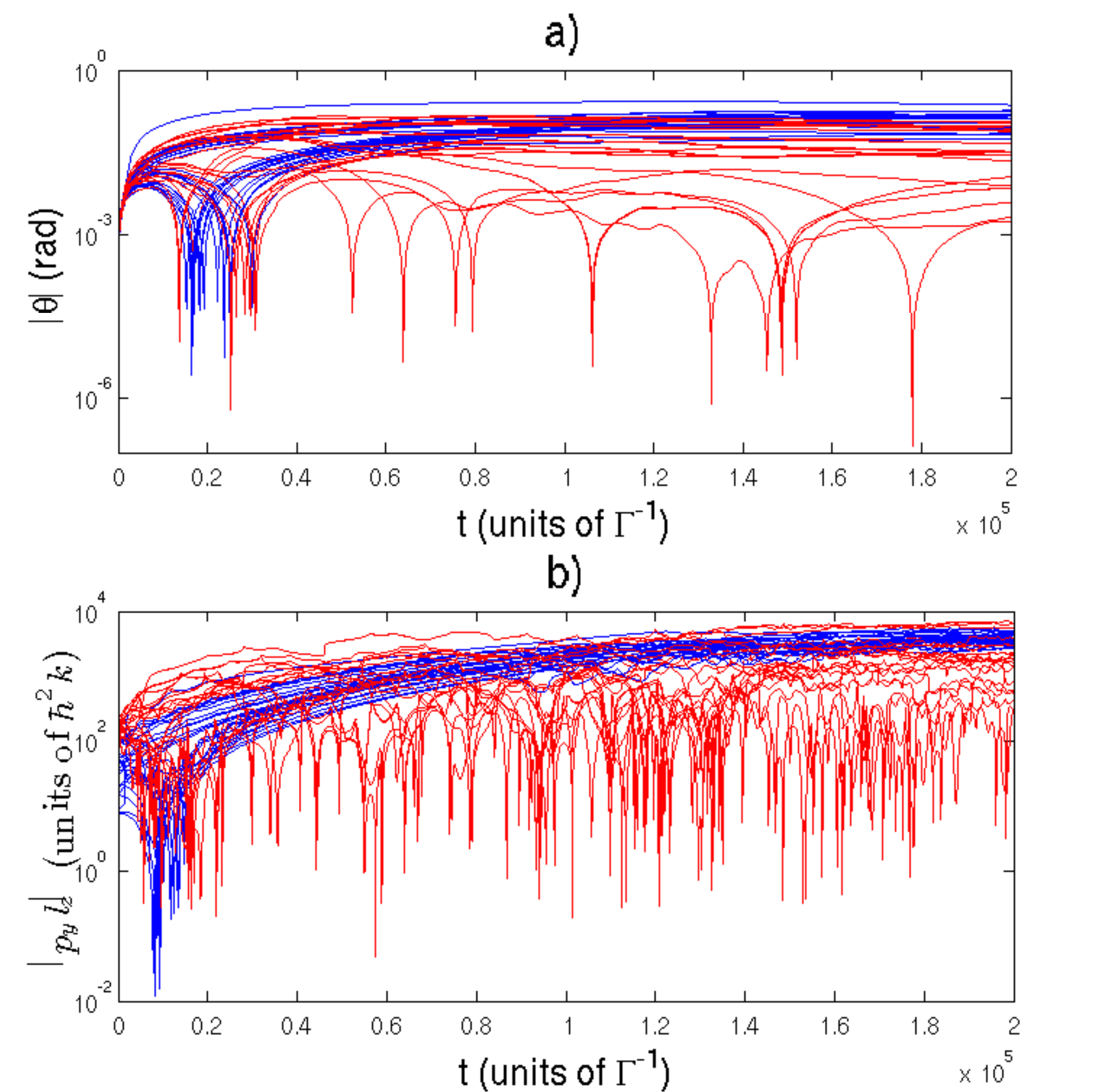}
\caption{\label{fig:Fig4}(Color Online) (a) Absolute value of the deviation angle, $\theta = \arctan{y/x}$ and (b) temporal evolution for the values of the mechanical property $\mathbb{A}_{atomic}=L_{z}P_{y}$  for different components of the clouds presented in Figs.~(\ref{fig:Fig3}a-b)}
\end{figure}

These results are illustrated in Fig.~(\ref{fig:Fig3}) for  $a=-5$ and $x_0 = 150\lambda$. If $mv_x^2/2k_B \sim 1.5\mu$K and the irradiance is $1.725~ W/ cm^2$, the atoms that arrive to the beam within a distance of $\sim \lambda$ to the $x$-axis are violently deflected, while those arriving in a zone $\lambda<\vert y\vert< 2u_M^2\sim 50\lambda$ have  deflection angles up to $\theta_d^{max}\sim$0.17 radians, Fig.~(\ref{fig:Fig3}a); those arriving at $80\lambda<\vert y\vert< 150\lambda$ move almost vertically, Fig.~(\ref{fig:Fig3}b). If the irradiance is almost quadrupled the deflection angle reach values of the order $\theta_d^{max}\sim$0.35 radians for atoms arriving at the dark zone, Fig.~(\ref{fig:Fig3}c), while the atoms arriving at $80\lambda<\vert y\vert< 120\lambda$ are partially focused towards the $x$-axis, Fig.~(\ref{fig:Fig3}d). This effect is also interesting by itself since it could be used for merging independent clouds. In Fig.~(\ref{fig:Fig4}a), the deflection angle, $\theta_d$, is illustrated for all arriving zones. Typical values of $\theta_d$ are tenths of
radians for atoms arriving at the dark zone of the beam.

Due to the parabolic symmetry of the problem, the product  of the atomic angular momentum $L_z$ and the linear momentum component $P_y$, $\mathbb{A}_{atomic}=L_zP_y$, is a relevant dynamical property for the description of the atomic cloud. As mentioned above, Weber beams carry a well defined value of the similar  EM property $\mathbb{A}_{EM}$, Eq.~(\ref{eq:A}). Through its classical motion, each atom  locally detects the density $d\mathbb{A}_{EM} /dV$, and could modify its $\mathbb{A}_{atomic}$ value. For the examples illustrated in Fig.~(\ref{fig:Fig3}a,b), initially the atoms have  $\mathbb{A}_{atomic}(0)\leq 150\hbar^2k_\perp$ and finally highly deflected atoms have $10^3\hbar^2k_\perp<\mathbb{A}_{atomic}<10^4\hbar^2k_\perp$ as shown in Fig.~(\ref{fig:Fig4}b). A detailed analysis lets us see that  atoms arriving outside the dark beam region may exhibit high fluctuations in their $\mathbb{A}_{atomic}$ value as a function of time, while atoms focused into the same bright parabola for $x<0$ have similar asymptotic $\mathbb{A}_{atomic}$ values. These results show
that Weber photons may transfer the dynamical variable $\mathbb{A}$ to atoms.

\begin{figure}[ht!]
\includegraphics[width= 0.9 \textwidth]{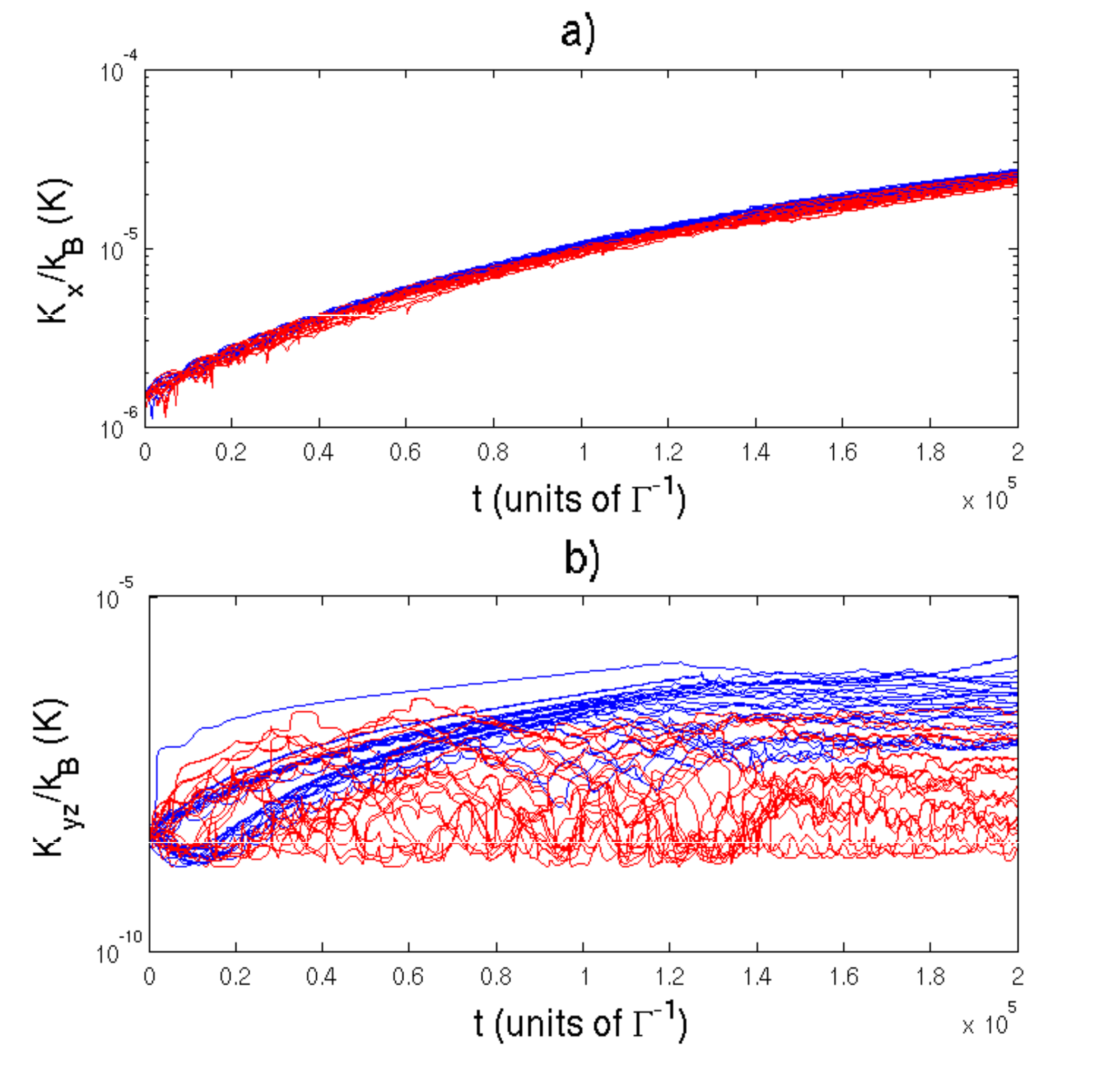}
\caption{\label{fig:Fig5}(Color Online) Contribution to the kinetic energy from, (a)  velocities parallel to the gravitational force direction, $K_x/k_B =m v^2_x / 2 k_{B}$, and (b) velocities perpendicular to the gravitational force direction, $K_{yz}/k_B=m v^2_\perp / 2 k_{B}$, for different components of the clouds presented in Figs.~(\ref{fig:Fig3}a-b) }
\end{figure}

As for the kinetic energy acquired by the atoms, we divide it in two parts: the contribution due to the velocity in the gravitational direction $K_x=mv_x^2/2$ and that in the transverse direction $K_{yz} = mv_\perp^2/2$. For the example illustrated in Fig.~(\ref{fig:Fig5}a), initially $K_x(0)/k_B\sim 1.5\mu$K and $K_{yz}(0)/k_B\sim 0.02\mu$K finally  $K_x/k_B\sim 20\mu$K and $K_{yz}/k_B(0)< 1\mu$K for atoms arriving at the dark region, and  $K_{yz}(0)/k_B\sim 1\mu$K for atoms arriving in the other regions. The movement in the $x$-direction is mainly driven by the gravitational field. The spreading in the kinetic energies is smaller for atoms arriving at the dark zone. At short times, $0<t<1.5\times10^4 \Gamma$, $\vert v_x\vert$ diminishes while  in the average $v_\perp\sim \vert v_y\vert$ increases. That is, the atomic deflection takes place mainly in that time interval. Thus, if smaller final kinetic energy for the atoms is desired, the region of the optical field could be shortened along the $x$-axis once the atomic thermal cloud has been split.

Summarizing, we have presented a scheme to split a cold non-interacting atomic cloud using a
single Weber beam. This scheme requires no further optics or electronics but
that used to generate the Weber beam. The atomic deflection angles are a function of the Weber parameter $a$ and the irradiance of the beam, as well as the initial kinetic energy and arriving zone of the atoms. We have shown numerically that it is possible to manipulate the population of the split components of the atomic cloud through the control of the entry point of the cloud with respect to the center of symmetry of the EM field intensity pattern. In fact, we have shown that this arrival dependence may be used to convert the light beam into a merging device. The atoms, when guided by a
parabolic bright region of the beam, acquire values of  $\mathbb{A}_{atomic}=L_zP_y$
with small dispersion. The main source of kinetic energy for the atoms is provided by the gravitational field.

For atoms that initially have kinetic energies in the quantum regime, lower irradiance of the light beam is expected to achieve similar deflection angles as those reported here. Under such conditions, studies on the quantum evolution in phase space of the atomic beam and its consequences on interferometric experiments would be especially interesting due to the topological structure of the EM Weber field \cite{us}.


\begin{thebibliography}{99}

\bibitem{Houde2000} O. Houde, D. Kadio, and L. Pruvost, {\it Phys. Rev. Lett.} {\bf 85}, 5543-5546 (2000).


\bibitem{Bragg1988} P.~J. Martin et al.,{\it Phys. Rev. Lett.} {\textbf 60}, 515 (1988); D.~M. Giltner, R.~W. McGowan, and S.~A. Lee, {\it Phys. Rev. A} {\textbf 52}, 3966 (1995); M.~K. Oberthaler et al., {\it Phys. Rev. Lett.} {\textbf 77}, 4980 (1996); S. Kunze, S. D\"urr, and G. Rempe, {\it Europhys. Lett.} {\bf 34}, 343,(1996); M. Kozuma et al., {\it Phys. Rev. Lett.} {\bf  82}, 871 (1999).

\bibitem{single}F. Riehle et al., {\it Phys. Rev. Lett.} {\textbf 67}, 177 (1991); D.~S. Weiss, B.~C. Young, and S. Chu, {\it Phys. Rev. Lett.} {\textbf 70}, 2706 (1993); T. Pfau et al., {\it Phys. Rev. Lett.} {\textbf 71}, 3427 (1993); M. Weitz, B.~C. Young, and S. Chu, {\it Phys. Rev. Lett.} {\textbf 73}, 2563 (1994); M. Weitz, T. Heupel, and T.W. H\"ansch, {\it Phys. Rev. Lett.} \textbf {77}, 2356 (1996).


\bibitem{Wang2005} Y.~J. Wang et. al. {\it Phys. Rev. Lett.} \textbf{94}, 090405 (2005).

\bibitem{Pasquini2005} T.~A. Pasquini et. al. {\it J Phys: Conf. Series} \textbf{19}, 139-145 (2005).

\bibitem{Kraft2005} S. Kraft {\it et. al}, {\it Eur. Phys. J. D} \textbf{35}, 119 (2005).

\bibitem{Hommelhoff2005} P. Hommelhoff et. al. {\it New J. Phys.} {\bf 7}, 3 (2005).

\bibitem{Dholakia} K. Dholakia and W.~M. Lee, {\it Adv. in Atomic Mol. and Opt. Phys.} {\bf 56}, 261-337 (2008).

\bibitem{OAM_BEC}  M.~F. Andersen et. al. {\it Phys. Rev. Lett.} \textbf{97}, 170406 (2006); R. Pugatch et. al. {\it Phys. Rev. Lett.} \textbf{98} 203601 (2007); D. Moretti, D. Felinto, and J.~W.~R. Tabosa {\it Phys. Rev. A} \textbf{79}, 023825 (2009).

\bibitem{allen92}  L. Allen, M.~W. Beijersbergen, R.~J.~C. Spreeuw, and J.~P. Woerdman, {\it Phys. Rev. A} \textbf{45}, 8185 (1992).

\bibitem{TabosaP99}  J.~W.~R. Tabosa and D.~V. Petrov,  {\it Phys. Rev. Lett.} \textbf{83}, 4967 (1999).

\bibitem{luciana}  M.~Babiker et. al. {\it Phys. Rev. Lett.} \textbf{89}, 143601 (2002).

\bibitem{bec current}  G.~S. Paraoanu, {\it Phys. Rev. A} \textbf{67}, 023607 (2003).

\bibitem{nienhuis}  H.~L. Haroutyunyan, G. Nienhuis, {\it Phys. Rev. A} \textbf{70}, 063408 (2004).

\bibitem{twist}  M. Bhattacharya, {\it Opt. Commun.} \textbf{279}, 219 (2007).

\bibitem{loops08} K. Volke-Sep\'ulveda and R. J\'auregui,{\it J. Phys. B: At. Mol. Opt. Phys.} {\bf 42}, 085303 (2009).

\bibitem{blas08} B.~M. Rodr\'{\i}guez-Lara and R. J\'auregui, {\it Phys Rev. A} {\bf 78}, 033813 (2008).

\bibitem{Stratton1941} J.~A. Stratton, {\it Electromagnetic Theory} (McGraw-Hill, New York 1941).

\bibitem{Lebedev1972} N.~N. Lebedev, {\it Special Functions and their Applications} (Dover Publications, U. S. A., 1972).

\bibitem{Abramowitz1972} M. Abramowitz and I.~A. Stegun, {\it Handbook of Mathematical Functions} (Dover Publications, U. S. A., 1972).

\bibitem{Volke2006} K. Volke-Sep\'ulveda and E. Ley-Koo, {\it  J. Opt. A: Pure Appl. Opt.} {\bf 8}, 867 (2006).

\bibitem{Bandres2004} M. Bandres et. al. {\it Opt. Lett.} {\bf 29}, 44-46 (2004).

\bibitem{us} B.~M. Rodr\'{\i}guez-Lara and R. J\'{a}uregui, {\it Phys. Rev. A} {\bf 79}, 055806 (2009).


\bibitem{Lopez2005} C. L\'opez-Mariscal, M.~A. Bandres, and J.~C. Guti\'errez-Vega, {\it Opt. Exp.} {\bf 13}, 2364-1369 (2005).

\bibitem{Durnin} J. Durnin, J.~J. Miceli Jr., and J.~H. Eberly, {\it Phys. Rev. Lett.} {\bf 58}, 1499 (1987).

\bibitem{arrizon} R.~Ponce-D\'{\i}az et. al. {\it Proc. SPIE} {\bf 6311}, 63110D (2006).

\bibitem{Gordon1980}  J.~P. Gordon and A. Ashkin, {\it  Phys. Rev. A} \textbf{21}, 1606 (1980).

\bibitem{Miller1993}  J.~D. Miller, R.~A. Cline, and D.~J. Heinzen, {\it Phys. Rev. A} \textbf{47}, R4567 (1993).

\end{thebibliography}
\end{document}